%% file: main.tex
\documentclass[acmsmall]{acmart}
\usepackage{array}

\usepackage{color, colortbl}	
\definecolor{Gray}{gray}{0.9}
\definecolor{LightCyan}{rgb}{0.88,1,1}

\newcolumntype{L}{>{\centering\arraybackslash}m{3cm}}
\AtBeginDocument{%
  \providecommand\BibTeX{{%
    \normalfont B\kern-0.5em{\scshape i\kern-0.25em b}\kern-0.8em\TeX}}}

\acmBooktitle{}

\newcommand{\yang}[1]{{\color{red} \textbf{(Yang: #1)}}}
\newcommand{\tanusree}[1]{{\color{cyan} \textbf{(Tanusree: #1)}}}
\newcommand{\yun}[1]{{\color{blue} \textbf{(Yun: #1)}}}
\newcommand{\zhixuan}[1]{{\color{brown} \textbf{(Zhixuan: #1)}}}
\newcommand{\review}[1]{{\color{red} \textbf{(Review: #1)}}}
\newcommand{\update}[1]{{\color{black}#1}}
\begin{document}

\title[Unpacking Creators' Practices with Non-Fungible Tokens (NFTs) and Their Communities]{``It's A Blessing and A Curse'': 
Unpacking Creators' Practices with Non-Fungible Tokens (NFTs) and Their Communities}

\author{Tanusree Sharma$^{*}$}
\email{tsharma6@illinois.edu}
\affiliation{%
 \institution{University of Illinois at Urbana-Champaign}
 \city{Champaign}
 \country{USA}
}

\author{Zhixuan Zhou$^{*}$}
\thanks{*The first two authors contribute equally to this paper.}
\email{zz78@illinois.edu}
\affiliation{%
  \institution{University of Illinois at Urbana-Champaign}
  \city{Champaign}
  \country{USA}
}

\author{Yun Huang}
\email{yunhuang@illinois.edu}
\affiliation{%
  \institution{University of Illinois at Urbana-Champaign}
  \city{Champaign}
  \country{USA}
}

\author{Yang Wang}
\email{yvw@illinois.edu}
\affiliation{%
  \institution{University of Illinois at Urbana-Champaign}
  \city{Champaign}
  \country{USA}
}

\renewcommand{\shortauthors}{Sharma and Zhou, et al.}

\begin{figure}[b!]
\label{fig:NFT}
\centering
\includegraphics[width=0.8\textwidth]{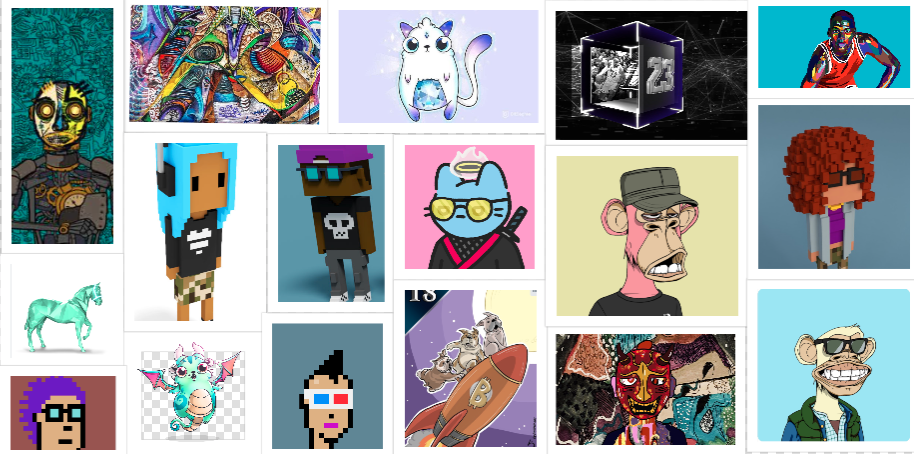}
\caption{Sample popular NFTs, e.g., CryptoPunks, Bored Ape, Meebits, Cool Cat, ZED RUN, Hashmask, and NBA Top Shots. Some of them are worth millions of dollars, e.g., CryptoPunks (volume: 6,596 ETH). 
As of the time of writing, 1 ETH is worth around \$3,300 USD.} 
\end{figure}

\begin{abstract}

 NFTs (Non-Fungible Tokens) are blockchain-based cryptographic tokens to represent ownership of unique content such as images, videos, or 3D objects. 
 Despite NFTs' increasing popularity and skyrocketing trading prices, little is known about people's perceptions of and experiences with NFTs. In this work, we focus on NFT creators and present results of an exploratory qualitative study in which we interviewed 15 NFT creators from nine different countries. Our participants had nuanced feelings about NFTs and their communities. 
 We found that most of our participants were enthusiastic about the underlying technologies and how they empower individuals to express their creativity and pursue new business models of content creation. Our participants also gave kudos to the NFT communities that have supported them to learn, collaborate, and grow in their NFT endeavors. However, these positivities were juxtaposed by their accounts of the many challenges that they encountered and thorny issues that the NFT ecosystem is grappling with around ownership of digital content, low-quality NFTs, scams, possible money laundering, and regulations.
 %
 We discuss how the built-in properties (e.g., decentralization) of blockchains and NFTs might have contributed to some of these issues. We present design implications on how to improve the NFT ecosystem (e.g., making NFTs even more accessible to newcomers and the broader population). 
\end{abstract}



\maketitle

\input{section/1_Introduction}

\input{section/2_Related-Work-2}

\input{section/3_Methodology}
\input{section/4_Findings-3}

\input{section/5_Discussion-2}
\input{section/6_Conclusion}

\bibliographystyle{ACM-Reference-Format}
\bibliography{sample-base}

\input{section/7_Appendix}

\end{document}

%% file: section/1_Introduction.tex
\section{Introduction}
\update{``Grimes sold \$6 million worth of digital art as NFTs.'' \cite{verge2} 
``What is an NFT? Are NFTs just a get-rich-quick scheme?''~\cite{guardian} ``Will NFTs transform the art world? Are they even art? ''~\cite{washinton}}

Non-Fungible Tokens (NFTs) are unique digital tokens that represent ownership of a particular piece of content such as digital art work, game objects, and collectibles. 
NFTs replicate the properties of physical items, such as scarcity, uniqueness and proof of ownership on a public blockchain (e.g., Ethereum) using cryptographic mechanisms and are usually traded in online marketplaces with cryptocurrencies \cite{nft}.
A blockchain generally refers to decentralized ledger systems that are append-only, i.e., once data has been written onto the blockchain, they cannot be altered \cite{wang2021non}. In the case of NFTs, once the NFT is created on a blockchain, one's ownership of the NFT underlying asset (e.g., an image) is public and can be viewed and verified by anyone.

Technically, an NFT can be thought of as an one-to-one mapping between a user account (i.e., a cryptocurrency wallet address) and the NFT's underlying asset or content which can be represented by a distinct Uniform Resource Identifier (URI). NFTs enable the establishment of the ``provenance'' of digital objects, offering information such as the original creator and owner history of a particular digital  object, and the number of copies of this object (a measure of scarcity). Typically, NFTs have their own special traits including provable scarcity, inter-operability and indivisibility, which make them attractive to creators, collectors and investors. 


%

\begin{figure}[t!]
\centering
\includegraphics[width=0.9\textwidth]{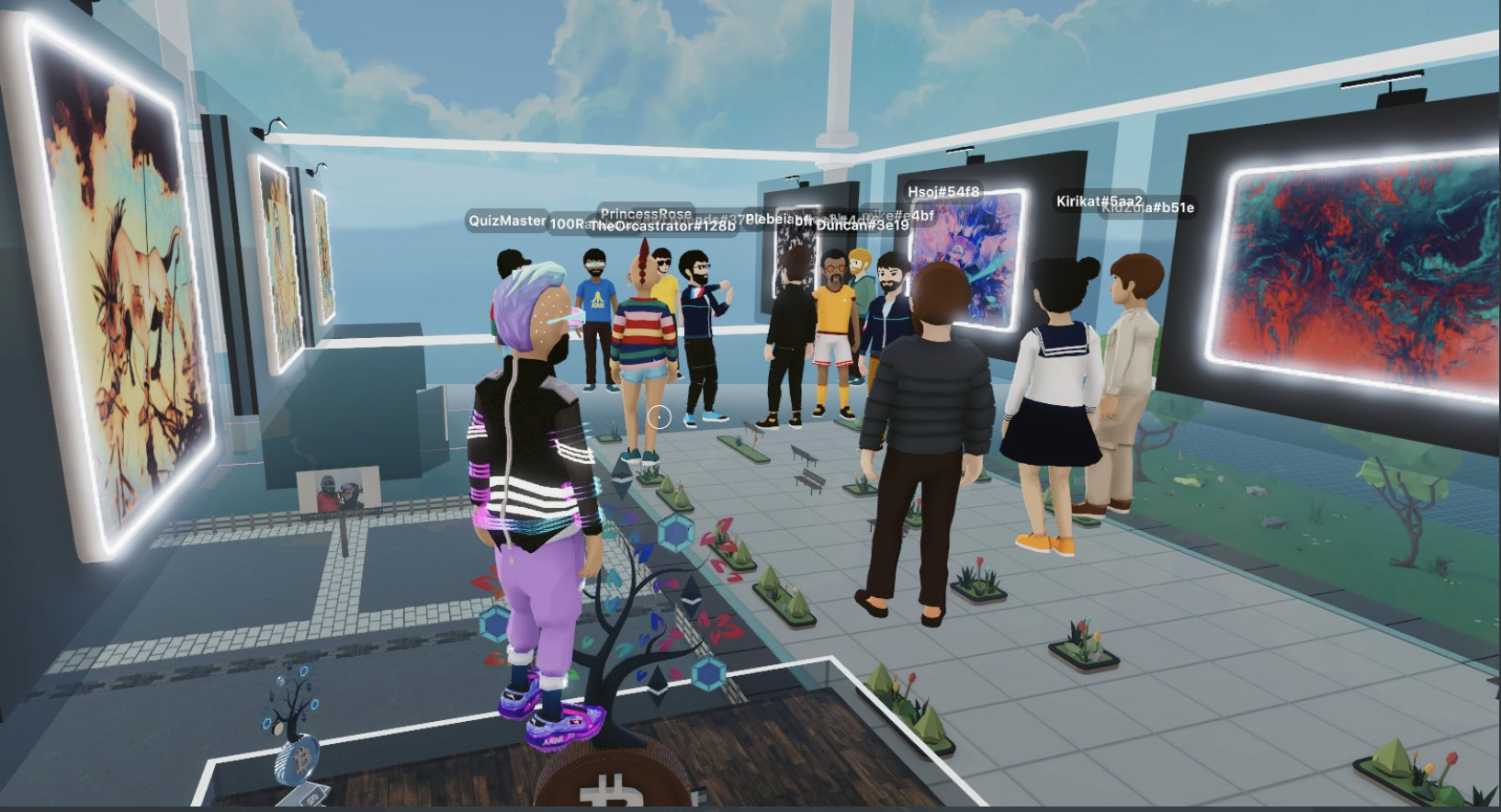}
\caption{A virtual NFT gallery in Decentraland, owned by P5. Galleries are interactive media in the game, designed for people to sell and buy NFTs. Each user has her own avatar/character in the virtual world. 
}~\label{tab:gallery}
\end{figure}

NFTs have gain lots of momentum in recent years. Early NFT projects such as CryptoKitties and CyberPunks are images released in 2017 and have sold for millions of dollars. 
NFTs in the gaming industry (i.e., trade of in-game objects) have already reached a certain level of maturity \cite{nadini2021mapping}, and many other industries such as music, sports and fashion are also experimenting with this emerging technology \cite{nft, million}. 
For instance, Decentraland is an online/VR game (more recently known as an example of metaverse) for creating and trading virtual land, on which users can run their own business, e.g., running an NFT gallery (see Fig~\ref{tab:gallery}). 
NFTs of famous sports players are getting sold for hundreds of thousands of dollars \cite{BBC}. Companies such as Adidas created their own NFTs.
Music producers (e.g., Grimes)
earned millions of dollars by selling NFTs of their recordings \cite{BBC1}. 



Another typical use case of NFTs is art. A telling example is that a digital artist known as Beeple sold a NFT of a digital drawing for a jaw-dropping price of \$69 million dollars~\cite{VERGE}. 
While counterfeits have been a long-standing problem in the traditional art industry, 
NFT with its ability of validating ownership and authenticity was thought as a possible solution \cite{BBC}. 
People can create and sell NFTs on online marketplaces, such as OpenSea and Rarible. 
Many believe that NFTs are poised to transform the art world, changing not only how art is bought and sold, but also, what kind of art people value. 

Despite the growing popularity of NFTs, extant research on this emerging phenomenon  is scant, mostly focusing on the technical aspects of blockchain-based protocols \cite{wang2021non} and market trade networks \cite{nadini2021mapping}. 
NFT is an inherently socio-technical system and we know little about people's perceptions of and experiences with NFTs. 
To help fill this research gap, we conducted semi-structured interviews with 15 NFT creators from nine different countries in 2021. 
Specifically, we sought to answer the following research questions: 

\begin{itemize}
\item RQ1: Why do NFT creators create NFTs? 

\item RQ2: How do NFT creators engage with NFTs and their communities?  
\item RQ3: What challenges do NFT creators encounter? 
\end{itemize}

Our results paint a rather nuanced picture of the NFT ecosystem. 
Most of our participants were enthusiastic about NFTs as a revolutionary technology that democratizes content creation and commercialization. Some even called it the ``New Renaissance.'' Furthermore, they felt that the NFT communities have played an integral role in their NFT experiences, supporting them to learn about, create, promote, and sell NFTs. However, their passion was also juxtaposed with various challenges they have encountered (e.g., learning and using new concepts and technologies) as well as their accounts of the ``dark side'' (e.g., proliferation of low-quality NFTs, scams and even possible money laundering) that the NFT ecosystem is grappling with. 





Our work makes three main contributions: (1) our study provides novel insights into  how NFT creators perceive and engage with NFTs, including their participation in the NFT communities as well as the challenges they have encountered (e.g., minting NFTs, assessing NFT quality); (2) our study identifies a few key issues that the NFT ecosystem faces (e.g., content ownership in practice, and regulations); and (3) we offer concrete design implications for improving the NFT ecosystems (e.g., making NFTs more accessible to the broader population).

%% file: section/2_Related-Work-2.tex
\section{Related Work}
In this section, we present literature on the history of NFTs, and the underlying blockchain ecosystem. We then compare the interactions and dynamics in the NFT marketplace/community to those in the traditional art market/community.

\subsection{The Rise of NFTs, \update{and the Blockchain Ecosystem}} 
\label{sec:blockchain}
\noindent \textbf{NFTs.}
NFTs are digital assets that represent the ownership of objects ranging from image and audio files, email and Twitter accounts, to art, gaming and sports collectibles \cite{abarbanel2019vgo}, which are encoded into the blockchain with smart contracts, and traded with cryptocurrency such as USDT and ETH \cite{evans2019cryptokitties}. Through NFTs, a unit of data is stored on a blockchain that certifies a digital asset to be unique and therefore not interchangeable as fungible tokens such as Bitcoin. Unlike other digital assets on blockchain, NFTs have unique certification of ownership, and are catching more attention for their uniqueness and provable scarcity recently. Notably, NFTs have enabled the verification of ``provenance'' of underlying assets, by validating such information as who owns a certain items, who previously owned or created that item, as well as how many copies exist for that item \cite{evans2019cryptokitties}. 

Originally, NFTs were based on the Ethereum blockchain, but increasingly more blockchains have implemented their own versions of NFTs \cite{wood2014ethereum}. Ethereum provides a secure environment for executing smart contracts, which are computer programs to carry out predefined functions. Ethereum token standards guarantee that a certain asset will behave in a specific way, and specify how people can interact with basic functions of the asset. 
Several Ethereum standards support the creation and trading of NFTs. For instance, ERC-721 \cite{721} proposed back in 2018 is exclusively used to create NFTs. It provides such functionalities as transferring tokens from one account to another, getting the current token balance of an account, getting the owner of a specific token, and getting the total supply of a token available on the network \cite{721}. ERC-1155 \cite{1155} proposed in the same year can also be used to create NFTs. There are also blockchain customized for NFTs, e.g., Flow \cite{onflow}.

NFT creators and buyers are the two main stakeholders in the market. A creator would create an NFT by signing a transaction, including the hash of the NFT data, and sending the transaction to the underlying smart contract. When the smart contract receives the transaction with NFT data, the minting process begins. The minted NFT is associated with a unique identifier in the blockchain, which is an evidence of ownership for NFT creators/owners \cite{hong2019design}. NFT marketplaces allow users to ``bid'' for NFT collectibles where buyers need to place a bid to meet the reserve price set by the creators, outbid other buyers, and finally settle the price to win\cite{hong2019design}. When tradings of a certain NFT are confirmed, the NFT metadata and ownership details are added to a new block to preserve the ownership chain. Creators can also ``burn'' their NFTs to completely erase art pieces from the blockchain. The entire process is built upon the underlying blockchain which stores information electronically in the distributed ledger.

\noindent \textbf{The Underlying Blockchain.}
Blockchain was proposed by Satoshi Nakamoto, where a proof of work (PoW) mechanism was used to reach a consensus on transaction data among untrusted participants in a peer-to-peer network \cite{nakamoto2019bitcoin, garay2017bitcoin} to ensure the integrity of the entire blockchain.
In addition to transaction data, each block contains a timestamp, the hash value of the previous block  \cite{chuen2017handbook}, by which fraud can be effectively prevented, since changes of a block would immediately change the respective hash value. 

Bitcoin \cite{bitcoin}, the first application of blockchain, is a cryptocurrency where owners have full control over the currency, and are able to transact it at their own discretion and without geographical constraints. It excludes involvement of any centralized authorities such as governments and banks, and is free of regulation and censorship, thus protecting anonymity of users. 

The spectrum of blockchain applications ranges from financial services (e.g., decentralized exchanges, settlement of financial assets, and payment systems), Internet of Things, to public and social services \cite{zheng2018blockchain, yli2016current}. NFTs are among the examples where blockchain is applied to a setting beyond cryptocurrency \cite{conti2018survey}. 

There is a small but emerging body of research in HCI concerning blockchain, largely the experiences, motivations and values of Bitcoin users. For example, Gao et al. indicated the misconception of crypto non-users of being incapable of using Bitcoin, since they did not fully understand how the protocol functions and if it satisfies properties of usual payment systems they currently use \cite{gao2016two}. Lustig studied online communities of Bitcoin to explore users' values and trust in the algorithmic protocol \cite{lustig2015algorithmic}, showing that users preferred to integrate human judgment to mediate algorithmic authority for fairness. Kow et al.  described the cultural affinities of early crypto adopters as a sociocultural journey \cite{kow2016hey}. This study indicated that meaningful community building and cultural enterprise, in addition to providing technical support, are important for early stage users  \cite{kow2016hey}.

\noindent \textbf{History of NFT Projects.}
The notion of NFTs had become popular before NFT standards were established on Ethereum \cite{NFT-history}. Back in 2017, memes started to be traded on Ethereum. In the same year, John Watkinson and Matt Hall created unique characters, i.e., Cryptopunks, on the Ethereum blockchain. Each character was different, and the number of characters was limited to 10,000. The first project using the ERC-721 NFT standard was the famous CryptoKitties, which was on every news station, from CoinDesk to CNN. People made enormous profits by trading the virtual cats. The years of 2018 and 2019 saw a massive growth within the NFT ecosystem. There were 6.1 million trades of 4.7 million NFTs, primarily on Ethereum and WAX blockchains, between June 23, 2017 and April 27, 2021 \cite{nadini2021mapping}. 
Hundreds of NFT projects have emerged in the past few years. NFT marketplaces were also thriving, led by OpenSea, Rarible, Foundation, etc. Web3 wallets such as MetaMask further made it easier for people to participate in the NFT space. Public attention towards NFTs exploded in 2021, when the market experienced many record-breaking sales. One such example is the digital artist known as Beeple, who sold an NFT of his work for \$69 million \cite{VERGE}. That sale positioned him ``among the top three most valuable living artists,'' according to the auction house. Another popular NFT project in 2021 was Bored Ape Yacht Club, which offered ten thousand unique iterations of the cartoon primates for sale, each at the price of about two hundred dollars. The initial batch of NFTs were sold in more than two million dollars, and eventually increased to fourteen thousand dollars for each item in trading value \cite{NEW-YORKER}. Yet another big sale was ``meetbit'' NFT collection by Larva Labs. ``Meebits'' is a series of 20,000 distinctive 3D avatars designed to be used in the metaverse, which was initially priced at 2.5 ETH per item \cite{cryptobriefing}.

\noindent \textbf{Advantages and Challenges of Blockchain and NFTs.} 
NFT projects (e.g., games and collectibles), infrastructures (e.g., crypto wallets), and NFT marketplaces (e.g., Opensea and Rarible) are built upon the underlying blockchain. The decentralization feature of the blockchain enables crypto applications to be tamper-resistant \cite{crime}. However, some unwanted features also come with decentralization, e.g., lack of scalability \cite{scalability}, high gas fees, and lack of regulation.

Gas fees on the Ethereum blockchain is especially high, given the high volume of on-chain traffic. 
On Ethereum, users who pay higher fees will have their transactions confirmed faster. The high gas fees is one of the barriers for NFT creators to mint and trade their work. 
Furthermore, lack of regulation in the blockchain has enabled malicious users to involve in crimes, e.g., money laundering, and trading illegal items \cite{crime}. Since blockchain hides the association between a transaction and the person who conducts it, it is hard to trace the illegal transactions. 

NFTs as a new technology built on blockchain inherit some of the limitations mentioned above. Thus it is important to examine both NFT-specific and blockchain-induced drawbacks, toward creating a better NFT landscape.

\subsection{\textcolor{black}{NFT Marketplaces and Communities}}


\noindent \textbf{Market Structures, Stakeholders, and Characteristics.}
The NFT marketplace has some intrinsic similarities with traditional art exhibitions and auctions. They mainly have two parties: creators/owner/seller and buyers. In traditional art markets, objects offered for sale are unique pieces, the value of which is determined by their characteristics (e.g., topic, medium, and reputation of artists), subjective non-monetary utility \cite{goetzmann2011art}, resale value, and demand (e.g., wealth concentration, equity market evolution, changes in art-collecting audience) \cite{lovo2018model}. Stakeholders interact through different social interactions, such as visual and material conduct of bidding and buying, during trading \cite{auction}.
In contrast, NFTs are collections of art, music, games, etc., the trading of which is supported by the underlying distributed ledger, with consistent records of transactions. NFT tokens are unique and can have different value depending on its rarity, quality, and reputation of creators. NFT marketplaces allow users to bid for NFT collectibles. When an NFT is sold or resold, its metadata and ownership details are added to a new block on the blockchain, thus the ownership history is recorded and preserved.

The main difference between the two marketplaces is that the ownership chain is not well structured in traditional art markets and may lead to fraudulent events. Whereas NFT tokens are created and traded on top of blockchain, the ownership of which can not be easily tampered. From literature, we find that information on provenance, such as details about an art object's pedigree, is important, as it may increase people's confidence in the authenticity of the art object and help identify forgery \cite{van2002art}. How to handle the issue of forgery in the traditional art space has always been an open problem in previous literature. 
NFTs provide such a feature backed by the underlying blockchain, bridging the gap of traceability which may improve its trustworthiness.

\noindent \textbf{Societal Benefits of Creative Markets.} 
Art exhibitions 
provide a platform for artists around the world, and support their careers with creative development, exhibition and publication opportunities. For the audience, exhibitions are an environment to learn about art in an interactive manner, and socialize with other artists and visitors \cite{vosinakis2011virtual}. Research has shown evidence of positive relationship between the engagement in art and physical/psychological well-being of people \cite{o2012impact}. Arts also provide a paradigm of interaction to enhance people's experience, and support their engagement, learning, and alignment with community \cite{rizzo2007fire}. There have been attempts for virtual art exhibitions to allow local and remote visitors to co-exist in the environment, access the interactive content, and communicate with each other \cite{vosinakis2011virtual}. 

There is a similar trend in the NFT marketplace,
where artists create different types of collectibles and arts, and exhibit/sell them virtually .
However, there has been a lack of research on the emerging, virtual NFT space. We aim to bridge this research gap by understanding the benefits of NFTs for people who engage in the NFT marketplace, with an emphasis on creators. 

\noindent \textbf{Motivations and Dynamics of Community Engagement.}
People in a group, either in the physical world or social media space, may share similar interest in art, politics, sports, etc. Research shows that socializing, as a vital social activity, is not necessarily task-related; people also socialize with other individuals for emotional reasons \cite{community}. People with similar interests tend to support certain events together, e.g., posting and cheering for certain sports teams, using slogans and videos \cite{fandom}. Different ways can facilitate social interaction, such as image sharing, and text-based communication \cite{interaction}. Even everyday objects, augmented with technologies, can contribute to disseminating people's lifestyle values \cite{cheon2021jarvis}. The circulation of values may inform us about the alternative roles of technologies in speaking for our norms in a community environment \cite{cheon2019beg}. In a previous study on online innovation communities \cite{innovation}, intrinsic motivations prove to be effective for initial participation and long term engagement . It is shown that the majority of participants are engaged to the online innovation community for fun, and activities that challenge them. In most cases, participants' roles are contributors and collaborators. Extrinsic motivations such as reward also attract people who are more interested in objective values \cite{lakhani2003hackers}.

In the NFT marketplace, there is a trend of forming active communities. There are many online communities and forums for NFT users, such as Discord channels of popular NFT projects, and NFT trading platforms (e.g., Rarible, Foundation, Opensea). Such gathering of individuals with a common interest provides NFT creators with an opportunity to share their accomplishments and values in public platforms. In this work, we seek to explore people's motivations for joining the NFT communities, which are innovation communities for artists. We also explore the community dynamics and interactions, toward understanding NFT practices from both individual and community perspectives.

%% file: section/3_Methodology.tex
\section{Methodology}

To answer our research questions, we conducted an exploratory qualitative study, in which we interviewed 15 NFT creators from different countries. 
To help get familiar with the emerging phenomena around NFTs, especially the community dynamics, we joined various NFT channels, forums, and communities on social platforms such as Discord, Telegram, Slack, and Twitter. Such observation has helped us build rapport with community members, and helped us recruit participants for interviews. 
Each interview was conducted online (e.g., via Zoom) in 2021 and took about 45 minutes. Each participant received a \$10 Amazon e-gift card. This study is IRB approved. 

\subsection{Participant Recruitment}
Our study focused on exploring NFT creators' perceptions, practices, and challenges.  
NFT creators are people who create various forms of NFTs, such as images, videos, music, game, and 3D models. They sell the NFTs they created in NFT marketplaces (e.g., Rarible, OpenSea), and may also purchase and collect others' NFTs, for instance, as a way of giving back to the NFT community. 

Participant recruitment was challenging in part because NFT creators are often very busy with their NFT work. To help reach NFT creators, we recruited using different methods: (1) posting recruitment information on Twitter and Discord channels, (2) snowball sampling, and (3) direct message to NFT creators who promoted their NFTs on Twitter. Prospective participants were invited to fill out a screening survey, which asked if they had experience with creating NFTs, and if they were 18 years or older. These were our recruitment criteria. In the end, we had a total of 15 participants. Seven participants were recruited from Twitter and Discord postings, five were from snowball sampling, and three were from direct messages. The participation in our study was completely voluntary, and participants were allowed to quit at any time. 

\subsection{Participant Demographics}
Our 15 participants had diverse backgrounds. They were from many NFT communities based on their area of expertise and interest, such as art, music, video games, and memes. However, all of them were in the broader NFT community on Reddit \cite{reddit}, Discord \cite{discord}, and Twitter.
Our participants were from nine different countries: United States, China, Iran, Canada, Germany, Belarus, Switzerland, Argentina, and United Arab Emirates.  
Our participants also had diverse occupations, including students, business professionals, software programmers, a tattoo artist, and a landscape photographer. Some of them were doing NFTs full-time. Their years of experience with NFTs ranged from less than one year to two years, at the time of the interviews. Most participants used at least one NFT marketplace (e.g., Rarible, Opensea, Foundation) to create and promote NFTs. 

However, our sample was heavily skewed in gender identities. 13 of them self-identified as male and only two as female. While this is not ideal, this skewed distribution might be reflective of the underlying gender disparities in the cryptocurrency space in general and the NFT communities more specifically, which recent reports seem to suggest. For instance, 
a recent study conducted by Art Tactic~\cite{arttactic} reported that 77\% of capital in NFT marketplace, earned by male artists \cite{coindesk}. 
A Pew Research report also suggests that in the US, the likelihood for a man to use cryptocurrencies is double of that of their female counterparts~\cite{pew}.

%
Table~\ref{tab:table1} summarizes the demographics and other backgrounds of our participants.

\noindent 

\begin{table*}[b!]
  \centering
   \caption{Basic information of study participants. For NFT creation, three major marketplaces were used. r: Rarible, os: Opensea, tz: Tezos. Price range is given in ETH (1ETH=\$3200USD at the time of writing). Note that gender is self-identified: M indicates male, and F indicates female, na indicates data that not available.
  }~\label{tab:table1}
  \resizebox{\columnwidth}{!}
  {\begin{tabular}{l r r r r r r r r r}
    \hline
    {\small\textit{ID}}
    & {\small \textit{Gender}}
    & {\small \textit{Country}}
    & {\small \textit{Role}}
    & {\small \textit{Year(s) of exp}}
    & {\small \textit{Main platform(s)}}
    & {\small \textit{Total created}}
    & {\small \textit{Price range(ETH)}}
    & {\small \textit{Specific art/media}}
    \\
    \midrule
    P1 & M & UAE & Creator & <1 & Rarible&28(r)&0.1-0.15&AI tools, Illustration\\
    P2 & M & Canada & Creator & <1 & Rarible&27(r)&0.003-0.05&3D, VR with Tiltbrush \\    
    P3 & M & USA & Creator & <1 & Twitter NFT group&108(r)&0.1-3&2D, 3D art\\
    P4 & F & Germany & Creator & <1 & Rarible & 98(r) & 0.05-0.1 & 3D ducks, 2D Linework\\
    P5 & M & USA & Creator & <1 & Rarible, Decentraland&29(r)&0.3-4&3D game, holographic\\
    P6 & M & Argentina & Creator & 1 & Opensea, Rarible&205(os) &0.005-5& 3D  abstract porn art and tattoo\\
    P7 & M & China & Creator+Collector & <1 & Opensea & na & na & na\\
    P8 & M & USA & Creator & 1 & Rarible & 160 & 0.013-5 & Piet Mondrian Vector Mash-Ups\\
    P9 & F & China & Creator & <1 & Own platform & na & na & na\\
    P10 & M & China & Creator+Collector & <1 & Own platform, Rarible & na & na & na\\
    P11 & M & Taiwan & Creator+Collector & 2 & Opensea, Foundation & na & na & na\\
    P12 & M & USA & Creator & 1 & Opensea, Rarible, Tezos & 66(tz) & 0.01-5 & Experimental 4D/AV\\
    P13 & M & Switzerland & Creator & <1 & Foundation, Opensea, NFTB & 75(os) & 0.35-1.5 & Luminescence, Hyperreal Landscape  \\
    P14 & M & Iran & Creator & 1 & Foundation, Rarible, Opensea& 205(os) & 0.14-2 & Geometric shapes of faces, sculptures \\
    P15 & M & Belarus & Creator & <1 & Rarible, Foundation, Opensea& 12(os)&0.09-0.5 & Natural things and digital apocalypse  \\
    \hline
  \end{tabular}}
 
\end{table*}

\subsection{Interview Questions} 
%
%
The interviews were semi-structured. Our interview protocol had three major  sections that correspond to the three research questions. 
We started by asking about their perceptions of and experiences with NFTs, such as how they first got to know NFTs, their initial experiences, and their motivations for creating NFTs, and their perceived pros and cons of NFTs. 
For instance, \textit{``How did you first learn about NFTs? What's your first direct experience with NFTs? Can you tell me briefly what kinds of NFTs you are creating? (For each NFT) What is it about? Can you remember the reasons for creating it?''} We also asked them about their experience in participating in the NFT communities. For instance, \textit{``Can you share any of your experiences that demotivate you from NFT engagement, or participating in the forums/online communities?''} All interview questions are listed in Appendix~\ref{interview}. 

The original interview script was in English. 
Since some participants from China did not speak English, a native Chinese speaking co-author translated the interview script into Chinese, and conducted the interviews in Mandarin Chinese. Other interviews were conducted in English. 

\subsection{Analysis}

The interviews were audio recorded from the Zoom meetings upon participants' permission. We transcribed the recordings, and analyzed the data using a thematic analysis~\cite{ta2020user}. 
Two researchers first read through the interview transcripts many times, and then proceeded to open code a small sample of transcripts independently. They met regularly to discuss and converged on a shared codebook before they continued to code other transcripts. Some examples codes are ``self-regulated business model,'' ``purchase NFTs,'' ``disruption of art,'' and ``difficulty in minting NFTs.''  
They then grouped relevant codes into emerging sub-themes and then further into main themes. Some example themes include NFTs empowers creators, mutual support in NFT communities, and challenges in learning new concepts and technologies related to NFTs. 
Since this work is exploratory, following the guidelines \cite{mcdonald2019reliability}, we did not calculate inter-coder reliability.
%
We did not learn anything significantly new from our last 2-3 interviewees, implying a sign of theoretical saturation. 
In this paper, we use participant quotes to illustrate the major themes. All quotes are anonymized.


%% file: section/4_Findings-3.tex
\section{Findings}

In this section, we present our participants' perceptions and practices of NFTs, specifically their motivations for creating NFTs (RQ1), 
their engagement with NFTs and the communities (RQ2), and the challenges they encountered regarding NFTs (RQ3).


\subsection{A Diverse Range of Motivators for NFT Creation (RQ1)}
\label{sec:motivation}
Our participants expressed various motivations for creating NFTs, such as exploring interesting properties of NFTs (i.e., uniqueness and proof of ownership); expressing personal feelings, emotions, pleasure and a sense of accomplishment in presenting creative artworks; leveraging new models for art business; and being a part of an innovative movement. Below we will elaborate on these different motivations for NFT engagement. 


\noindent \textbf{Exploring Interesting  Properties of NFTs.} 
Many participants (e.g., P1, P7, P8, P10) were drawn to NFTs because of their interesting properties such as uniqueness and proof of ownership. 

{\bf Uniqueness.} Each NFT is non-fungible and can have its unique traits, while fungible tokens (e.g., Bitcoin, Ether) are all the same. The uniqueness of NFTs makes them special and distinguishable from each other. People often associate personal identities, meanings or emotions with their NFTs, for instance, by using them as profile pictures (a.k.a., PFPs). 
P7 explained: 
\begin{quote}
    \textit{``Fungible tokens are more for speculation, and people trade them in the same way as stocks. But non-fungible tokens, most of them are kind of art. [...] They have some special traits, very different from each other. If you own one of them, you have some personal emotion for it, and would use it as an Avatar.'' 
    }
\end{quote}


{\bf Proof of Ownership.} 
Unlike traditional art, the authenticity, ownership, and provenance of which are challenging to verify~\cite{van2002art}, NFTs are valued for their traceability, rarity and ownership with the help of underlying blockchain technologies. NFTs allow people to check who (or more precisely, which wallet address) owns the underlying NFT content (e.g., image, video), because that information is public on-chain data. The underlying NFT content is usually saved in decentralized storage (e.g., IPFS) with a unique address. A screenshot or copy of the underlying NFT content, say an image, would yield a different piece of content (or NFT) with a different address. Each minted\footnote{``Minting" refers to the process of tokenizing an item, and uploading it to an NFT platform such as OpenSea.} item is encoded and recorded in the underlying public  blockchain~\cite{wang2021non}. P7 contrasted the pre-NFT and post-NFT era regarding authenticity confirmation, demonstrating the level of transparency that NFTs brought to the art market:
\begin{quote}
    \textit{``Before NFT was a thing, I would post paintings on Instagram, and offer them for sale. 99.9\% of people would just say, why should I buy it? I can just copy the image and download it for myself. There's no real way to prove that you created it or that you've bought it. Now every time it is sold, I get the percentage''}
\end{quote}
The ``percentage'' mentioned by P7 means that every time the NFT is (re)sold, the original creator could earn a portion of the sale price, which 
again demonstrates the traceability of NFTs. 


\noindent \textbf{Expressing Personal Emotions.} Expressing personal feelings and emotions was one of the key motivators for NFT creators (e.g., P1, P10, P11). 
Some participants mentioned creating NFTs, such as NBA Top Shot video clips to support their favorite sports teams and players. For example, P10 
offered emotional support to his favored sport team by creating NFT with the players' finest playing moments. In his case, NFT served to convey his attachment and support to the sports team. 

P11 shared another touching example. He did not have any technical background, but created NFTs for his son as a way to keep memories of him. 
He noted, 
\begin{quote}
    \textit{``I was thinking of creating NFT because my child will be born in December. I'm thinking of making some connections between the baby and the blockchain, and even create a trend on the market. I think I'm fine with Photoshop and simple art, as well as uploading it to the Ethereum blockchain. I want to do this because I want to keep some memories for him, like funny stories and his dreams.''}
\end{quote}
This case illustrates the emotional connection between real-world events and NFTs. 
Some participants also mentioned sending NFTs as postcards on special occasions to extend good wishes. 

 



\noindent \textbf{Expressing Creativity.} Many participants created NFTs to express their creative ideas (e.g., P2, P4, P5, P10, P13). These participants already had artistic background such as being landscape photographers, film maker and tattoo artists. 

Creating NFTs was seen by them as a way to express themselves, as they could create/mint NFTs out of their own ideas, rather than following clients' requirements. P4 was such an example: 
\begin{quote}
    \textit{``This was a way to step out of my comfort zone of normal art a little bit, start something new, learn and create something new. In my previous [tattoo] business, I created tattoo designs for my clients' ideas. But NFT are just my own ideas, and that's pretty cool to make something new.''}
\end{quote}
She viewed NFTs as a rich space for exploring her own artistic ideas. 
Similarly, P5 combined his multiple personal interests, such as writing, blogging, and video editing into NFT creation, which was seen by him as a way to express his talents. 



\noindent \textbf{Leveraging New Models for Art Business.} 
Many participants (e.g., P9, P11, P15) pointed out that NFTs empowered artists by providing them with a low barrier to present their artwork. P11 also added how the inclusive, supportive NFT marketplace enabled unknown and under-served artists to thrive. 
Unlike the traditional art market where artists often rely on event organizers, sponsors and vendors to exhibit and sell their art \cite{spence2020bob}, 
the decentralized blockchain ecosystem creates the possibility to remove intermediaries in the NFT marketplace. P13 indicated the NFT platform as an autonomous platform which empowered artists, who could exhibit, sell and promote their work out of their own will, and regulate the art work with the underlying smart contracts.


P3, who owned an NFT company, highlighted the convenience of the new business model for NFT creation due to the underlying blockchain. He thought NFT allowed freelancers to automate their workflow, with all transactions recorded by the blockchain and being easily exportable, which helped artists commercialize their art work more efficiently: 


\begin{quote}
    \textit{``It empowers individuals to create their own business model. Your business transactions would always be recorded [...] benefit freelancers to automate their workflow. You don't have to worry about tracking receipts. They're all stored on blockchain [...]in theory, if you use it right, if you trust the platform you work with, you would have unlimited control.''}
\end{quote}

The COVID-19 pandemic has also forced many artists to explore alternative business models because the lock-down policy shuts down their offline business. 
For instance, P13 was the inventor of the famous hyper-real landscape photography, and used to earn money through workshops and tutorials. The pandemic ceased his offline work, and he had to turn to the NFT market for a living: 
\begin{quote}
    \textit{``The events I used to attend for a living were almost not possible during the pandemic. So I lost almost 80\% of my income because of that. [...] I built a team with 2 guys. Now I try to sell as much NFT as possible in my spare time to earn enough money to pay my bills.''}
\end{quote}
P13 also created and donated NFTs to non-profit organizations, e.g., one that helps stray dogs. 

\update{We also noticed some participants with limited entrepreneurial/business/art opportunities started their careers in NFTs. For example, P4 was a female tattoo artist in Berlin who never got a chance to exhibit her artworks in physical exhibitions and who had to shut down business during the pandemic. She said, 
\begin{quote}
    \textit{``I never sell arts in the gallery in real life. I entered this NFT space three months ago and no one ever know me or see me [...] Now I have 3k followers in Twitter and my NFTs get noticed.''}
\end{quote}
The NFT space has made art work more accessible to her. Through her hard work and the help and promotion from the NFT community, her NFTs were gaining momentum. 
}

\noindent \textbf{Being A Part of ``The New Renaissance''.}
Many participants (e.g., P1, P3, P11, P12, P13) considered blockchain and NFTs revolutionary and they wanted to be a part of the movement. 
They called NFTs the new Renaissance for artists. P3 explained:
\begin{quote}
    \textit{``Almost everyone that I've interacted with believes this [NFT] to be one of the most important technologies and advancements in our lifetime. A lot of people are calling it the new Renaissance for artists[...]. A lot of new inspiration and new art is coming out, and people are very willing to collaborate in network, across countries, across the world.''}
\end{quote} 

P3 also noted the collaborative nature of NFT communities, which we will dive deeper in the next section on their actual engagements with NFTs and their communities.
It is also important to note that the empowerment of content creators by the NFT and underlying blockchain technologies could be exploited and misused. As we will see in Section~\ref{challenges}, our participants noted that the NFT space is also fraught with low-quality NFTs, scams and even the possibility of money laundering.

\subsection{Staged Engagement with NFTs and Their Communities (RQ2)}
\label{5.2}

%
In this section, we present our participants' actual practices with NFTs, including how they first learned about NFTs, how they created NFTs, and how they participated in the NFT communities. We highlight the staged engagement with NFTs, elaborating on different practices of novice and experienced NFT creators.  

\noindent \textbf{{First learning about NFTs.}} 
%
Our participants first learned about NFTs from different sources, such as family members, friends, and news media.  
For instance, P4's brother introduced NFTs to her as a way of selling pictures during the  pandemic lockdown when her shop was closed: 
\begin{quote}
   \textit{``Before I started with NFT, I didn't do crypto. My brother is investing in it for a longer time. During lockdown and pandemic we had to close our shops. [...] He told me about NFT and a new way to share, and sell the pictures to earn money. That was my entry in NFTs.''}
\end{quote}
Once she initially learned about NFTs, she started exploring and creating her own NFTs.

\begin{figure}[tb!]

\centering
\includegraphics[width=1.0\textwidth]{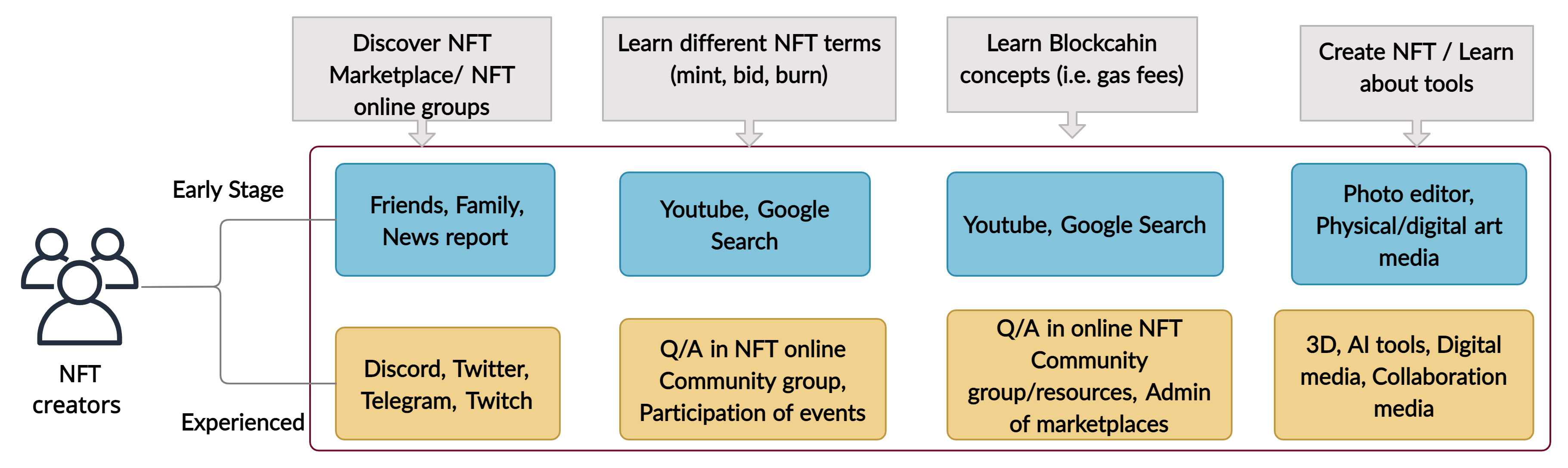}
\caption{Two stages of how our participants created NFTs:  an early stage (when they initially joined the NFT marketplace), and an  experienced stage (when they stayed in the marketplace for a few months). The top row shows the steps taken to create NFTs, whereas the bottom rows (blue and orange boxes) show the corresponding resources used for each step, for early stage and experienced stage, respectively. During the early stage, our participants tended to seek information and create NFTs on their own. In contrast, during the experienced stage, they tended to discover and participate in relevant online groups and communities, and create NFTs more collaboratively. 
%
%
}  
\label{process}
\end{figure}

\noindent \textbf{Creating NFTs.} \update{
To create NFTs, our participants reported mostly using NFT platforms or marketplaces (e.g., OpenSea, Rarible) rather than directly writing or using smart contracts.  
We also observed that as our participants learned more about NFTs and gained more experience with NFT marketplaces, their NFT creation practices tended to evolve. Therefore, we categorize their NFT creation practices into two different stages: (1) an early stage (i.e., when they initially joined the NFT marketplace), and (2) a more experienced stage (i.e., after they stayed in the marketplace for a few months).}
%
Figure~\ref{process} summarizes the NFT creation workflows and resources used for these two stages, respectively.

{\bf Early Stage.} During the early stage of NFT creation, most of our participants encountered difficulties due to lack of knowledge of underlying blockchain concepts, especially for those who did not have an IT background. This is similar to the traditional software development realm where new developers face challenges due to new terminologies and concepts \cite{aagerfalk2008benefits}. For P12, it took him a while to understand the new terminologies around NFTs and the associated crypto wallet. P6 felt the crypto world was changing fast and there was a lot to learn about cryptocurrencies in general:
\begin{quote}
    \textit{``When I started working with this community of artists, I didn't understand anything about NFT and the crypto world. It's moving really fast, so I've been studying a lot since last year. To create and mint NFT, it is important to know general cryptocurrency.''}
\end{quote} 
During the early stage, our participants often relied on themselves, for instance, by 
searching for resources in Google and/or YouTube with keywords such as ``create NFTs.'' Some participants were not aware that Ethereum gas fees were irreversible once the minting transaction was sent to the blockchain/validators, and thus lost their money by uploading incorrect NFTs to the marketplace.


\update{{\bf Experienced Stage.} NFT creators who have stayed in the marketplace for a few months and became more experienced consequently discovered different online NFT community groups on platforms such as Discord, Twitter, and Telegram. Instead of relying on themselves as in the early stage, they tended to participate in the NFT groups and seek help there, for instance, by directly asking NFT platform developers or other community members when they encountered bugs during the minting, creating, and promoting process. They may even send queries on art media and tools.} 

For instance, P3 had difficulty in setting up a wallet and depositing cryptocurrency into it to mint his NFT. He asked the questions on Twitter, and was directly replied by a platform representative:
\begin{quote}
    \textit{``It was actually kind of difficult to set up a wallet on that blockchain, and to deposit the cryptocurrency. I asked around and tweeted about my challenge. And then one of their representatives directly replied to me, and was able to help me.''}
\end{quote}

\update{Our participants mentioned Discord and Telegram groups as their primary venues for asking questions about and discussing NFTs. 
In general, they relied on NFT communities and online groups as useful resources for help,  rather than searching/learning NFT resources by themselves.

}



{\bf Creating Advanced NFTs.} In the experienced stage, our participants also learned to use advanced tools to create NFTs in various modalities, such as videos and 3D art. 
For instance, P1 disliked purely digital art and instead created NFTs with traditional paintings using AI tools. 
\begin{quote}
    \textit{``I've always been a fan of fine art that was made in the 1600s, 1700s, 1800s. I'm not a huge fan of the overly digital art that's very popular now. It looks a little bit childish.''}
\end{quote}
While P1 chose not to share the details of the AI tool he used, which he thought was ``private,'' he explained the final product, a GIF file combining 10 different stages of an evolutionary landscape: 
\begin{quote}
    \textit{``This is an evolutionary landscape, so it's not a still image. It's actually a video that contains 10 different stages of an evolution, evolving throughout time. Then I mint a GIF that represents what the painting actually looks like as a whole. This is actually a piece of art using AI tools.''}
\end{quote}

Unlike P1, many participants (e.g., P2, P3, P4, P5, P6, P12, P15) preferred purely digital NFTs. For instance, P4 created 3D digital duckies with different colors, textures and costumes using Blender, a 3D modeling tool.  
Some preferred NFTs with practical value, and may combine them with other cutting edge technologies. For instance, P10's team combined NFTs with VR:  
\begin{quote}
    \textit{``We created immersive VR scenarios such as an office, a cinema, or a pub. I can directly participate in some pop bar activities in VR. They're especially useful during the pandemic when you stay home every day. It's different from just sitting in front of your computer, because it requires you to interact. Not just a visual nature, it is more based on the experience of the entire environment ad interaction.''}
\end{quote} 
In the blockchain-based virtual world (a.k.a., ``metaverse''), virtual items are created, minted and traded as NFTs. P10 believed such NFTs had more value than other popular NFTs such as the NBA Top Shot NFTs which were ``merely for collection purposes.''

P2 created 3D NFTs using Tilt Brush \cite{tiltbrush}, a room-scale 3D-painting virtual-reality (VR) application as well as Photoshop and other VR software.
\begin{quote}
    \textit{``I've known NFT for five months, but started creation only recently, using Photoshop, Tilt Brush and VR. This is a crazy place to manipulate sketches into digital format.''}
\end{quote}


While some participants made their NFTs from scratch, others used their earlier artworks. For example, P13, a landscape photographer, created an NFT based on the pictures he took.

\noindent \textbf{Promoting NFTs.} After creating/minting NFTs, the creators need to promote the NFTs. Many participants 
talked about promoting NFTs via the official social media channels (e.g., Discord, Twitter) of the projects. 
For instance, P1 explained the typical use of Discord: 
 \begin{quote}
     \textit{``If you're the owner of a personal [NFT] project, you can talk to your community via Discord, engage with them, and answer any questions of potential buyers.''}
\end{quote} 

\update{P2 also used these Discord channels to learn about people's interests and preferences for NFTs. He then used these insights gained to create his next NFTs.} 
\begin{quote}
     \textit{``I'm in 15 Discord channels. There are members who are buyers. It's great to connect with people, knowing what type of work they like, so I can import those in my NFT.''}
\end{quote} 

In addition to social media channels, NFT projects also organized events to engage with community members. 
For instance, P5, an owner of an NFT company, organized online events to promote a new NFT game his company created. They engaged with their community members in Decentraland, and live streamed the events on Twitch \cite{twitch}. He explained: 
\begin{quote}
\textit{``We're throwing an event to open our gallery in Decentraland. 
[...] Now we stream on Twitch almost every day, spending hours in the game, meeting people, and then helping them learn how to play NFT games and create content. Creating relationships with the community and then building things together is a lot of fun.''} 
\end{quote}
Such social practices of NFTs P5 mentioned were considered very helpful by our participants in general, which we will detail next. 

\noindent\textbf{Collaboration and Mutual Support among NFT Creators.} Many participants highlighted the collaborative and supportive nature of the NFT communities. 
For instance, P6 explained: 
\begin{quote}
    \textit{``We're really collaborative and always trying to make valuable pieces of art, not trying to make products just for sale. We're always thinking about proposing something new.''}
\end{quote}

P12 identified himself as a ``community-based person,'' and enjoyed resource sharing and helping each other in the NFT community. 
%
Some participants mentioned showing support to new creators by buying their NFTs. P4 mentioned buying NFTs of a new artist just like other community members:
\begin{quote}
    \textit{``There's part of the community that will buy your first NFTs. They just want to get you started. Some think that you might become something really big, so they buy your first NFTs. And others are doing it like `hey, let's just get you started. Let's get you moving, get you excited, give you a little push.''}
\end{quote} 
This kind of community support is reciprocal. For instance, P1 regularly gave back to the community by buying others' NFTs whenever he reached a milestone in his own NFT project: 
\begin{quote}
    \textit{``I set up a goal in my mind: when I reach a milestone, I'll give back to the community by buying an NFT. And after I sold my first NFT, I bought one back.''}
\end{quote} 

Another supportive behavior among creators was following-back on Twitter to increase the fan base for each other.  
P4 elaborated on this community practice: 
\begin{quote}
    \textit{``I entered this space three months ago. I didn't have a Twitter profile before that. If you're open to just being a part of the community and growing with others, then you get a lot of followers and friends there to support you.''}
\end{quote}

However, the NFT communities are not always risk free. Next, we will present the major challenges faced by NFT creators, including the potential risks in the NFT communities.

\subsection{Challenges Encountered by NFT Creators (RQ3)}
\label{challenges}
Since NFTs are a relatively new phenomenon, our participants repeatedly described the challenges they encountered, such as the difficulty of learning new and complex NFT concepts and technologies, the lack of metrics to assess (new) NFT creators and NFTs, and the prevalence of phishing NFTs and scams. Table~\ref{norms} summarizes these challenges, as well as how they were addressed by our participants.

\begin{table*}[!t]
    \centering
    \caption{Challenges faced by NFT creators and their current strategies to address these challenges.}

    \small
    \resizebox{13cm}{!}{
    \begin{tabular}{p{.2\linewidth}|p{.8\linewidth}}
  \textbf{Challenges} & \textbf{Current strategies to address the challenges}\\
    \midrule
       Difficulty in understanding new and complex concepts \& technologies & \begin{itemize}
  \item Seek help from community members, e.g., sharing tools/resources, solving bugs
  \item Seek help from NFT project/platform administrators upon request
  \item Search answers online by themselves (e.g., Google, YouTube)
  \end{itemize} \\
    \midrule
     Lack of assessment metrics for (new) NFT creators \& NFTs & \begin{itemize}
  \item Look at prior reputation and how well-known the creator is
  \item Look at the number of their social media followers (mainly Twitter accounts)
  \item Estimate the amount of work/time devoted to creating the NFT
  \item Assess aesthetics of the NFT
  \end{itemize} \\
      \midrule
    Prevalence of phishing NFTs and scams & \begin{itemize}
  \item Manually inspect NFTs and compare them with existing NFTs
  \item Report scams to NFT project administrators
  \item Alert/Remind other community members (mostly via Discord channels and Twitter)
  \item (Some NFT platforms) Use a KYC-like process and/or an ``invite-only'' mechanism to restrict community membership
  \end{itemize} \\
            \bottomrule
    \end{tabular}
    }
    \label{norms}
  \end{table*}

\noindent \textbf{Coping with New and Complex Concepts and Technologies.} \update{
Despite support from the community, many participants (e.g., P1, P2, P3, P4, P5, P11, P12, P15) repeatedly emphasized the challenges of learning and understanding the sophisticated concepts and terminologies of blockchain technologies especially during the early stage. 
For instance, P12 found connecting a NFT account with an existing wallet quite intimidating, especially when he encountered bugs during that process. He mentioned watching YouTube videos, which were far from instructional for lay people. 
\begin{quote}
    \textit{``In my opinion, NFT is still pretty tech heavy.[..] They're almost speaking another language and the system is complicated. They don't have guidelines in Rarible or other places to follow. So I've watched YouTube videos [...], not instructional at all [..]. It gets a little hazy when I'm going to put some money in it.''}
\end{quote}
P4 added that NFT marketplaces should have well-prepared lists of resources and tutorials for new creators to learn about concepts, tools, and strategies to improve/promote their NFTs. 

While the built-in append-only / irreversible property of the blockchain technologies (e.g., Ethereum) prevents malicious parties from tampering the data stored on chain, this feature was foreign to inexperienced NFT creators who might have been used to systems that support undos. 

\begin{quote}
    \textit{``You're preparing for an NFT, and you mint it out. Then two hours later, you realize you've made a typo, [...] the cover isn't the exact cover you wanted. It's set in a stone, so you can't edit it. Let's say you accidentally burned an NFT, then there's no way could it be restored.''}
\end{quote} 

The lack of undos costed them gas fees if they made mistakes in creating their expected NFTs. They not only struggled with grasping technical concepts but also explaining the new (desirable) business models enabled by NFTs. For instance, P2 had a hard time explaining NFT-related business ideas to older-generation business partners. He elaborated: 
\begin{quote}
    \textit{``You're in university, and you must be in your early 20s. The idea of owning a digital good in a video game is a natural thing for you. But for my generation, and especially the generation above me, it's like a foreign language. The number one thing that comes to my mind is making it less intimidating.''}
\end{quote} 

Besides understanding the technical and business concepts about NFTs, NFT creators still face thorny challenges such as assessing the quality of NFTs and detecting NFT scams. 
}

\noindent\textbf{Lack of Assessment Metrics for (New) NFT Creators and NFTs}.  
As we discussed in Section~\ref{sec:motivation}, the underlying decentralized blockchain technologies significantly lower the barriers for artists or others to create and sell their NFTs, who follow their own business models without the presence of intermediaries. However, this democratization of NFTs is currently coupled with a lack of standard or agreed-upon mechanisms to assess the quality of NFT creators and NFTs, which has in part given rise to the proliferation of low-quality NFTs. 
For instance, P1 shared his observation: 
\begin{quote}
    \textit{``They are just extremely desperate for people to see their work, and most of their art isn't actually that great. It's something that you'd make within 20 minutes on Photoshop. Unfortunately, that's because the gates are open and they are not accountable for anything.''}
\end{quote}
P1 used the estimated amount of work time needed as a quick proxy for quality while other mentioned proxies were prior reputation, and the number of followers on social media. Similar to P1, several other participants (e.g., P5, P8, P13, 15) felt low-quality NFTs demotivated NFT creators who put much work into their NFTs. For example, P13 felt discouraged by the fact that lower-quality NFTs sold better than his work.  He also pointed out that there were no mechanisms in the NFT marketplace to assess good creators and NFTs: 
\begin{quote}
    \textit{``I sold, since last month, nothing. That's a little bit demotivated if I see other guys are selling, even if their NFTs are very poor quality.''}
\end{quote} 

This lack of assessment metrics is particularly challenging for new creators because they might have just started out and do not yet have a reputation. Another related factor is that the pseudonymity of (new) creators pose additional challenges for trust. 
For instance, P8 mentioned that many NFT creators use pseudonyms, and there is a trend to use avatars as NFT creators' profiles. He personally felt uncomfortable communicating with pseudonymous peers who asked him for help, or sharing his NFT experience without knowing who he was talking to: 
\begin{quote}
  \textit{``Some new people use pen names for safety. It might be okay because the NFT platform allows it. But the phenomenon of avatar and not knowing who is behind it makes me trust less. Once a guy asked for help to create a piece, but I was not comfortable. You don't know who is behind." 
}    
\end{quote}

While the decentralized mantra of blockchains and NFTs empower people to pursue creative content creation and commercialization, the lack of government regulations and platform rules also contributes to the problem of low-quality NFTs. 
P1 eloquently put it: 
\begin{quote}
    \textit{``It's a blessing and a curse that there's not much regulation, because very bad art can be posted there, especially the gasless platforms have very, very, very saturated with very low quality and very low effort NFTs, and also there are legal concerns to people that might think that NFTs are used to launder money. And I do personally think some NFTs are being used to launder money like the ones going through ridiculous prices.
    ''}
\end{quote}
By ``regulations'' P1 was referring to NFT platform or marketplace rules. P1 did go on to say having more regulations would also create barriers that make NFTs less accessible to the broader population. Our participants also reported not knowing any governmental regulations around NFTs (neither did us). We also do not know whether money laundering is actually happening via  NFTs, but these challenges of pseudonymity, trust, and regulations are even heightened when the NFT communities are sometimes fraught with phishing NFTs and scams that tried to trick people. 

\noindent\textbf{Difficulties of Detecting Phishing NFTs and Scams.}
While our participants noted the overall positive and supportive culture of these NFT communities, several participants (P2, P3, P6, P7, P8, P10, P11, P15) warned that malicious users often posted phishing NFTs and scams in the Discord and/or Twitter of mainstream NFT projects. They also noted the challenge of detecting such phishing NFTs and scams. One type of phishing NFTs mentioned by our participants was that malicious creators copied the content of others' existing NFTs to create their own NFTs, and then set a lower price than the original ones. Since buyers are often not aware of this fact, they tend to buy NFTs from these cunning creators due to cheaper prices. The copied content was sometimes modified with very minor changes that make them almost visually  indistinguishable from the original NFTs. While this kind of phishing NFTs do not seem to trick people to provide their sensitive information as traditional phishing attacks, we consider them  phishing because they pretend to be the original NFTs and mislead people to buy them instead. P7 explained: 
\begin{quote}
    \textit{``People can copy an image, add some new features with photoshop tools, and mint it on Opensea. Then showing people that this is the original one, but actually it's altered by some random people. And unfortunately, there is no way the system can identify this.''}
\end{quote}
While the NFT communities often crowdsourced and alerted their members about recent and potential NFT scams, this practice alone falls short. Some participants (e.g., P2, P11) mentioned they knew people who had fallen for this kind of scams and had big financial losses. 
Part of the challenge is that it is not always easy to quickly detect these phishing NFTs. While there are technical mechanisms (e.g., visual hashing) that can quickly assess the similarity of two images, we are not aware of any existing systems that have this feature in the domain of NFTs, as this participant suggested. However, this type of phishing NFTs should not be confused with legitimate derivative NFTs based on existing ones. For instance, Mutant Apes NFTs are derivatives of the Bored Apes NFTs. Derivative NFTs make it very clear that they are creative alteration and continuation of the original NFTs rather than pretending to be the original ones.

Recently, some NFT platforms started to explore ways to mitigate scam NFTs. For instance, according to P2, Rarible allowed new users to optionally verify their Twitter accounts in  the on-boarding process, which is similar to the KYC (Know Your Customer) process in traditional centralized platforms. However, it was not mandatory.
P11 mentioned that Foundation adopted another mechanism, i.e., invite-only, to restrict random users. P11 felt more confident in minting NFTs on Foundation compared to other platforms. He explained: 
\begin{quote}
    \textit{``If you want to join, you've got to have someone invite you.[...] When the platform just started, there were only several known artists. Only those invited by them could join, and could further invite 2 or 3 other artist, like a snowball. So I feel assured about the quality of NFTs on Foundation.''}
\end{quote} 
However, the invite-only mechanism may achieve its quality control at the cost of the inclusiveness of NFTs, which was one of the promising premises of NFTs and the underlying decentralized blockchain technologies. Some participants were disapproved of this mechanism. For example, P5 thought it as a steep barrier for new creators to enter the NFT space. He was also concerned that the invitation process could be biased.

%% file: section/5_Discussion-2.tex
\section{Discussion} 

Through the lens of our participants (who were NFT creators), our results paint a rather nuanced picture of the NFT ecosystem, which includes but not limited to: the underlying blockchain technologies with their built-in properties (e.g., decentralized, trustless); NFT platforms or marketplaces (e.g., OpenSea, Rariable); NFT projects (e.g., CryptoPunks, Bored Apes) and their online groups or communities (e.g., Discord, Twitter); NFT creators, sellers and collectors; as well as malicious actors (e.g., scammers). This is mostly a NFT creator's perspective of the NFT space. 


Our participants have expressed a mixed bag of feelings, emotions and tensions about the NFT ecosystem, ranging from the excitement about the underlying blockchain/NFT technologies, the empowerment of new forms of content creation and commercialization, and the appreciation for the community support and reciprocity, to the annoyance at the proliferation of low-quality NFTs, the contempt for scams, and the ambivalence about  regulations. Below, we will try to unpack these inter-related ideas emerged from the interviews of our participants.



\subsection{Is NFT The ``New Renaissance''?} 
Most participants were passionate about NFTs and some of them called it the ``New Renaissance.'' We will start with the positive sentiments and then follow up with the negative aspects. 

\noindent {\bf NFTs Support Creativity.}
Unlike fungible tokens such as Bitcoins, non-fungible tokens (NFTs) are unique, i.e., any two NFTs are different. This uniqueness allows content creators to fully explore their creativity in the digital realm. As we saw in the results section, one of the main motivations for NFT creation is indeed creators embedding  their personal identities, feelings and emotions into their NFTs. Our participants had different backgrounds, such as a blogger, a video editor, a tattoo artist, and a landscape photographer. In their traditional offline work, their creativity was often constrained by their client requirements. They saw NFTs as an exciting opportunity where they could express their own artistic ideas. 

\noindent {\bf NFTs Democratize Content Creation and Commercialization.}
Blockchains are the technical bedrock of NFTs.  
One of the key properties of blockchain technologies is decentralization, which often means storage, computation and even governance of the blockchain systems are decentralized, i.e., there is no centralized authority to make and enforce rules \cite{jahani2018scamcoins}. Critically, this decentralization can remove intermediaries in the marketplace. In the traditional art industry, there are often intermediaries (e.g., exhibit organizers, auction companies, artist agencies) that artists rely on~\cite{van2002art}. 

In comparison, the blockchain-enabled NFT ecosystem can remove many of these intermediaries and empower traditional artists and others to create and commercialize their digital content in form of NFTs. Many participants raved about how NFTs allow them to try new business models for their creative work in a more efficient and autonomous way. For instance, it is much easier for them to promote and sell their work as well as do the business accounting (all the transaction records are public and on-chain). Another recurring and encouraging theme is how NFTs empower many under-resourced artists and creative professionals, such as those who were from developing countries, who were less known, and who were in a poor financial standing during the COVID19 pandemic. Some of our participants (e.g., a tattoo artist and a photographer) were struggling to make a living during the pandemic due to the physical lockdown. Their venture into the NFT space has made a huge positive difference for them. 

NFTs not only revitalize traditional artists, they also lower the barriers of digital art for some participants who did not have an artistic background before. However, NFT still has a steep learning curve and could be further improved to make it more accessible to a broader population. We will revisit the accessibility improvements in Section~\ref{design}. 

\noindent {\bf NFT Communities Provide Mutual Support and Help  Collaborations.} 
Our participants also attributed part of their positive experiences with NFTs to the supportive NFT communities (e.g., official Discord channels of NFT projects). 
During the early stage of their NFT expeditions, they tended to take a {\em self- and independent-learning} approach where they learned the primary concepts around NFTs, such as, minting, bidding, burning as well as blockchain concepts, such as gas fees for minting NFTs from Google search and Youtube. In comparison, during the more experienced stage, they tended to follow a {\em social learning} approach, in which they relied on different NFT community groups in Discord, Twitter for advanced learning by Q/A and community events. In this stage, they played with their skills and learned to use different tools, such as Adobe Photoshop, Tiltbrush, VR software to create 3D arts, GIF, and sketch manipulation. Our participants mentioned NFT official project groups in Discord and Twitter as their main venue to promote their work. 
Research has shown that sustainability of organizations depend on how they are designed as social learning systems which built upon individuals' personal experience and sense of belonging through engagement and alignment \cite{wenger2010communities}.
Many participants commented on how the NFT communities have played an important role for their NFT promotion and collaborations.
Many also mentioned showing support to new creators by buying their first NFTs. 

The collaboration of NFT creation can be more transparent than that of tradition art creation, where the contribution of each individual artists to a collective art piece often remains vague. NFTs address this issue by using the underlying blockchain and smart contracts to record each artist's contribution. 
Some of our participants talked about their experience in collaborative NFT creation. POAP.art\cite{poap} was one example mentioned by one  participant, where people could collectively finish an NFT painting on the virtual canvas. 
Online meetings were also organized by NFT platform founders and developers to discuss how to create a better NFT ecosystem for artists. 
Literature has shown that the success of an innovation community can be measured in terms of the number of completed collaborations \cite{luther2010works}. Collaboration can also help foster a supportive and sustainable marketplace. Thus the NFT communities could further support collective and collaborative creation of NFTs as an opportunity to enhance the ties among NFT artists, and create a sustainable community. 
Though several participants expressed their interest in collaboration in the near future, mainstream NFT platforms such as Opensea and Rarible did not support such a function. Future research can explore the adoption of collaboration mechanisms in NFT marketplaces.

Our participants also repeatedly mentioned common community practices, such as alerting other community members about suspicious activities, scams, for instance.  
Our participants described some of the indicators they used to identify malicious actors: a) malicious actors hardly communicate with peers personally, b) they tend to frequently spam links of NFTs in different groups without any specific details of their content, and c) they randomly create contents which are inconsistent in the types of art. 
%
Previous literature on community practices explained how a collective understanding, accountability and mutual engagement, and shared communal resources (language, routines, style, sensibility) could build a meaningful community \cite{wenger2010communities}. 
The kind of community alerting practices we observed in our study is a good sign for the sustainability of the community. 

Next, we will dive deeper into the problematic aspects of NFT ecosystem, such as the prevalence of low-quality NFTs, scams and even the possibility of money laundering. 

\subsection{Tensions, Dilemmas and Challenges}

\noindent {\bf Is NFT True Ownership of Content?}
One dilemma that surfaced in our study was that -- does NFT actually mean ownership of content in practice? 
One might argue that Web 2.0 with centralized platforms such as YouTube has already unleashed the power of user-generated content. However, the content creators (e.g., YouTubers) do not truly own the content nor can they determine whether or how the content they created will be displayed on these centralized platforms. In contrast, a common  narrative of Web 3.0 where NFTs is a prime example thereof is that content creators own their content in the decentralized blockchain world. Does NFT actually equate to true ownership of the content in practice? Let's first recap the technical underpinnings of NFTs. 

The underlying blockchain protocols (e.g., ERC-721) provide a technical mechanism to create a data blob (basically an NFT) \cite{bal2019nftracer} that associates a unique (wallet) address with a unique piece of digital content that is usually stored in a decentralized storage (e.g., IPFS) \cite{benet2014ipfs}. This one-to-one mapping record is then stored permanently and publicly on a blockchain such as Ethereum (after the NFT  minting transaction is validated and confirmed) \cite{wang2021non}. Once this record is stored on-chain, it cannot be altered because blokchains are append-only. 
This technical mechanism of NFTs are supposed to provide the authenticity of content and the proof of ownership. But, does this really mean ownership for all practical purposes? 

Our interview results suggest that the blockchain technical mechanism alone does not solve the ownership question completely in practice. Many participants were very vocal about their loathing for the wide spread of low-quality NFTs and even scams. The decentralization of blockchain and NFT technologies lower the barriers of content creation but also opens the floodgate of NFTs.  
NFT technologies can support rehashing or creating derivative NFTs based on existing ones (derivative NFTs actually allow the owner of the  original NFTs to make a cut from the proceeds of the trading of derivative NFTs). However, NFT technologies also allow the creation of what we called ``phishing NFTs'' where malicious actors essentially make a copy (e.g., a screenshot) of the underlying content of a popular (and expensive) NFT, apply negligible changes to the copied content, and then create a new NFT and offer to sell it at a lower price than the original NFT, albeit pretending to be the original NFT and hoping to lure the unsophisticated buyers.  
This issue calls into question the intellectual property rights of digital content in NFTs. This naturally leads to our next question. 

\noindent {\bf Are Regulations Needed?} 
Regulation is another dilemma emerged from our results. Here regulations cover both government regulations and NFT platform rules. Our participants were mostly ambivalent about them. On one hand, they felt that the whole premise of NFTs and blockchains is decentralization, which is the antithesis of centralized regulations. They believed that having regulations would increase the barriers of entry for NFTs. On the other hand, they feared that the lack of regulations would harbor misuses of NFTs, which lead to low-quality NFTs, scams and even possible money laundering. Countries around the world are expediting their legislation around cryptocurrencies \cite{hendrickson2016political}. For instance, countries such as China have already banned crypto mining operations and cryptocurrency trading \cite{reuters}. The US government is also actively looking to regulate the crypto industry \cite{hendrickson2016political}. All of these legislative efforts will have a ripple effect on NFTs.


On the NFT platform / marketplace front, some of them have already started experimenting different types of restrictions to prevent scams or low-quality NFTs. For example, Foundation uses an ``invite-only" policy:  only invited artists can mint NFTs on the platform. Such efforts are intrinsically double-edged swords. Despite their potential effectiveness in improving NFT quality, they are essentially creating a private club, preventing talented yet under-resourced content creators from creating their own NFTs on the platform. This type of policies are hurdles for  creating an open-access domain for everyone, which goes against the initial vision of blockchain. Future research is needed to strike a good balance between decentralization and NFT quality control. The rise of decentralized autonomous organizations (DAOs) could be a promising direction as they often codify and enforce (via voting and smart contracts) emerging community norms. 

\noindent {\bf Challenges Encountered by NFT Creators.}
Our participants described many challenges in creating NFTs which includes learning about blockchain and NFT functions to operate throughout the creation process. Early stage NFT creators found the blockchain concepts, such as, gas fees, irreversibility of blockchain intimidating, especially when they encountered bugs during the process. They also faced technical difficulties in setting up wallets and depositing cryptocurrencies in order to mint NFTs. The lack of resources in many NFT marketplaces made it more challenging for newcomers to learn about those concepts. 
They also wasted their cryptocurrencies (for gas fees) if they made mistakes in minting the NFTs as they expected. Early stage creators were less familiar with the purpose and irreversible nature of gas fees. Minting NFTs actually has made the gas fees on Ethereum prohibitively expensive from time to time in 2021. The scalability of Ethereum is a fundamental challenge and there are many Layer 2 solutions (e.g., opportunistic or zero-knowledge rollups) and alternative Layer 1 blockchains (e.g., Solana, Avalanche, WAX) that can reduce the gas fees of minting NFTs. 
There is also a lack of standard assessment metrics for the quality of NFTs. To assess the NFT quality, our participants used some heuristics, such as prior reputation of the creators, and the number of social media followers. These heuristics are particularly challenging for new creators who do not (yet) have a strong fan base and reputation. 
Next, we will discuss design implications based on our findings that could help address some of these challenges.

\subsection{Design Implications}
\label{design}

\noindent {\bf Making NFTs Even More Accessible.} 
Our study results show that the NFT ecosystem has empowered traditional artists and others to create digital content and mint them as NFTs. However, many participants noted that NFT has a steep learning curve especially for those who did not have an IT or digital art background. 
\update{
Our study indicates that newbie NFT creators are often intimidated by the  tech-heavy NFT concepts and terminologies. The process of minting NFTs also introduces several challenges, including creating a crypto wallet, depositing cryptocurrencies (e.g., ETH tokens) into the wallet, linking the crypto wallet to NFT marketplaces (e.g., OpenSea), and finally minting NFTs. As we can see, the minting process includes several elements in the blockchain ecosystem, i.e., cryptocurrencies, crypto wallets and NFT marketplaces. The usability of these components are under-studied in the literature. 
Educational materials dedicated to blockchain concepts and instructional guides can be integrated as built-in resources in NFT marketplaces to make it easier for non-technical users to create NFTs. Chatbots or conversational agents about NFTs can also be built to support new users. The  conversational logic of these chatbots can be designed based on commonly asked questions in NFT forums or communities.}

\noindent {\bf Assessing NFT Quality.} 
%
%
Measuring the quality of NFT creators and NFTs has remained as an open problem. 
%
One idea is a creator dashboard or a public creator profile page. Creators are incentivized to fill out the details, e.g., prior experience in NFTs and other content creation, and social media accounts. If anyone provides faked or inaccurate information, the community might catch it and basically ruin their reputation and prospect in the community. Interested buyers or collectors can then check the creator profile before purchasing the NFTs. 
Our participants suggested several metrics for assessing the quality of NFTs and NFT creators, such as the number of followers on social media and in NFT marketplaces, record of interpersonal communication in the marketplace, history of prior spam incidents, and NFT ownership record, which could be integrated into the profile. With such a public profile system, NFT creators can also look for potential collaborations. Furthermore, machine learning models could be trained, taking the information mentioned above as the input, to classify legitimate vs. scam NFTs.  Providing the model predictions to users can help them make more informed buying  decisions, and reduce the chance of falling for scams.  

\noindent {\bf Supporting Community Moderation.} 
%
%
\update{
%
We have already seen that low-quality NFTs and scams are prevalent problems in the NFT ecosystem. Some NFT marketplaces have started some restrictions. For instance, Rarible gave its users the option to go through a Twitter verification process, and those who do so may gain more visibility for their NFTs. In addition, our participants indicated that NFT communities would alert its members about any sketchy NFTs. However, this usually involves mostly manual checking of the NFTs in question, and is thus hard to scale and keep up with the rate at which new NFTs are minted. 

Scalability of moderation practices can be improved by AI technologies, which have been shown to play an active role in moderating toxic behaviors in games and online forums. Kou and Gui presented AI-led moderation in games, where community members collectively developed explanations of AI-led decisions, most of which were automated punishments \cite{AI:game}. Crossmod is a learning-based system \cite{AI:reddit}, which provides moderation recommendations for Reddit moderators by leveraging a large corpus of previous moderator decisions. P7, who developed a private NFT project, also mentioned automatic filtering of irrelevant NFTs using off-the-shelf AI. We believe that one way to use AIs here is to apply existing computer vision techniques (e.g., visual hashing) to automatically compute visual similarity of two NFTs. This assumes that there is a database of existing NFTs, which is not far-fetched because virtually all NFTs should be public and on-chain. NFT marketplaces can already do this because they have all the data about the NFTs on their platforms. When visual hashing detects highly similar NFTs, it can alert community members. }

\subsection{Limitations and Future Work} 

Our study explores an emerging phenomenon and has many limitations. 

First, our sample size is relatively small (N=15). However, recruiting for this population is difficult and our participants had very diverse backgrounds and NFT experiences. They created different types of NFTs (e.g., image, video, 3D) and participated in different NFT projects. 
Our study focused on the NFT creators' perspective, which is inevitably limited because the NFT ecosystem has other stakeholders (e.g., collectors) who might have different opinions. However, some of our participants were also NFT collectors and sellers in addition to creators. While there could be self-selection bias as in virtually any user research, the diversity of our sample and their views makes it less of a concern. 
As with exploratory qualitative work in general, we cannot claim that our findings are exhaustive or representative of the overall NFT creator population. However, our rich interview data shed light on the actual practices of some NFT creators. 
Future work could conduct large-scale surveys to obtain quantitative data  (e.g., the percentage of NFT creators who encountered scams).

Second, while our participants were from nine different countries, we did not have more than five participants from a single country. 
Therefore, we could not really examine NFT practices fom a cultural/national perspective, which would be a promising direction for  future work.

Third, our findings on community practices were from our interviewees' self-reported data. Future work could complement this data with an observational or ethnographic study. 

Last but not least, most of our participants (13/15) were male. This is a limitation of our study, however, it also resembles the gender disparity in the NFT community in particular~\cite{coindesk} and the cryptocurrency space more broadly~\cite{pew}. Future studies could purposefully oversample female NFT creators. In addition, future research could further examine the demographic distribution of NFT users, the reasons why certain groups are less represented in the NFT space, as well as create ways to make the NFT space more accessible to a much broader population. 

%% file: section/6_Conclusion.tex
\section{Conclusion}
Non-Fungible Tokens (NFTs) are making waves in the worlds of cryptocurrencies and digital art. The innovations behind NFTs and their jaw-dropping trading prices make them a popular topic in recent news. To explore the NFT space, we conducted an exploratory study where we interviewed 15 NFT creators from nine countries. Our study gave voice to these NFT creators' perspective of ``the good, the bad and the ugly'' of NFTs. As new NFT projects are popping up everyday, the issues and dilemmas our participants pointed out around low-quality NFTs, scams, ownership of content in practice, and regulations will continue to evolve. This fascinating and fast-changing NFT ecosystem is inherently socio-technical. We believe this is a timely and rich site for future research.




%% file: section/7_Appendix.tex
\appendix
\section*{Appendix}

\section{Interview Questions}
\label{interview}

\subsection{Perceptions of and Experiences with NFTs}
\begin{enumerate}
    \item Could you briefly introduce yourself (personal background, experience with cryptocurrency)?
    \item How do you identify yourself while interacting with NFT (creator/collector/trader)? What does NFT mean to you?
    \item How did you first learn about NFT? What's your 1st direct experience with NFT?
    \item How would you explain NFT to a lay person?
    \item Are you familiar with other cryptocurrency such as Bitcoin? What do you think is the difference between Bitcoin and NFT?
    \item Can you tell me briefly what kinds of NFT you are holding? (For each NFT) What is it about? Can you remember the reason for buying/creating it?
    \item Have you ever created NFT? Can you remember and explain why you created NFT for the very first time? Can you tell me more about it?
    \item Have you ever bought NFT? Can you remember and explain why you bought NFT for the very first time? Can you tell me about the most recent NFT you bought?
    \item What is your motivation behind engaging in the NFT marketplace?
    \item Is there any personal stories associated with the NFT you buy/create? Would you share one of your personal stories associated with NFT?
    \item Do you have any concerns or challenges in your experiences with NFT?
\end{enumerate}

\subsection{Community Perceptions and Practices}
\begin{enumerate}
    \item From your experience in this marketplace, can you explain how other community members are perceiving the value of NFT?
    \item Have you observed common topics, values and norms in online forums that you want to share with us?
    \item What do you think are the benefits of NFT marketplace and community?
\end{enumerate}

\subsection{Challenges, Risks, and Regulations}
\begin{enumerate}
    \item Can you share any of your experiences that demotivate you from NFT engagement, or participating in the forums/online communities?
    \item What do you think about regulation practices around the NFT marketplace?
    \item Can you also talk about your preferences and suggestions for regulation that you think is necessary in the NFT marketplace? 
\end{enumerate}